\documentclass[journal=jacsat,manuscript=article]{achemso}

\usepackage{bm}

\usepackage{mathrsfs}
\usepackage{amsmath, amssymb, amsfonts, amsthm,rotating,booktabs,array}
\usepackage{cancel}
\usepackage{bm}
\usepackage{booktabs}
\usepackage{subfigure}
\usepackage{graphicx}
\usepackage{mathabx}
\usepackage{setspace}
\usepackage{arydshln} 
\usepackage{algorithm,algpseudocode}
\usepackage{mathtools}
\usepackage{multicol}
\usepackage{multirow}
\usepackage{verbatim}
\usepackage{enumerate}
\usepackage{hyperref}
\usepackage{booktabs}
\usepackage{xcolor}
\usepackage{amsmath, amssymb}
\usepackage{nameref}

\usepackage{hyperref}
\usepackage{xr}

\author{Clayton Ellis}
\affiliation{Department of Statistics and Applied Probability, University of California, Santa Barbara, California 93106, United States}

\author{Xinyi Fang}
\affiliation{Department of Statistics and Applied Probability, University of California, Santa Barbara, California 93106, United States}

\author{Christopher Balzer}
\affiliation{Materials Research Laboratory, University of California, Santa Barbara, California 93106-5121, United States}

\author{Timothy Quah}
\affiliation{Department of Chemical Engineering, University of California, Santa Barbara, California 93106, United States}

\author{M. Scott Shell}
\affiliation{Department of Chemical Engineering, University of California, Santa Barbara, California 93106, United States}
\email{shell@engineering.ucsb.edu}

\author{Glenn H. Fredrickson}
\affiliation{Department of Chemical Engineering, University of California, Santa Barbara, California 93106, United States}
\alsoaffiliation{Materials Department, University of California, Santa Barbara, California 93106-5050, United States}
\alsoaffiliation{Materials Research Laboratory, University of California, Santa Barbara, California 93106-5121, United States}
\email{ghf@mrl.ucsb.edu}

\author{Mengyang Gu}
\affiliation{Department of Statistics and Applied Probability, University of California, Santa Barbara, California 93106, United States}
\email{mengyang@pstat.ucsb.edu}

\title[]
  {Fast  phase  prediction of charged polymer blends by white-box machine learning surrogates}
   
\abbreviations{}

\begin{document}


\begin{tocentry}




    \centering
    \includegraphics[width=8.3cm]{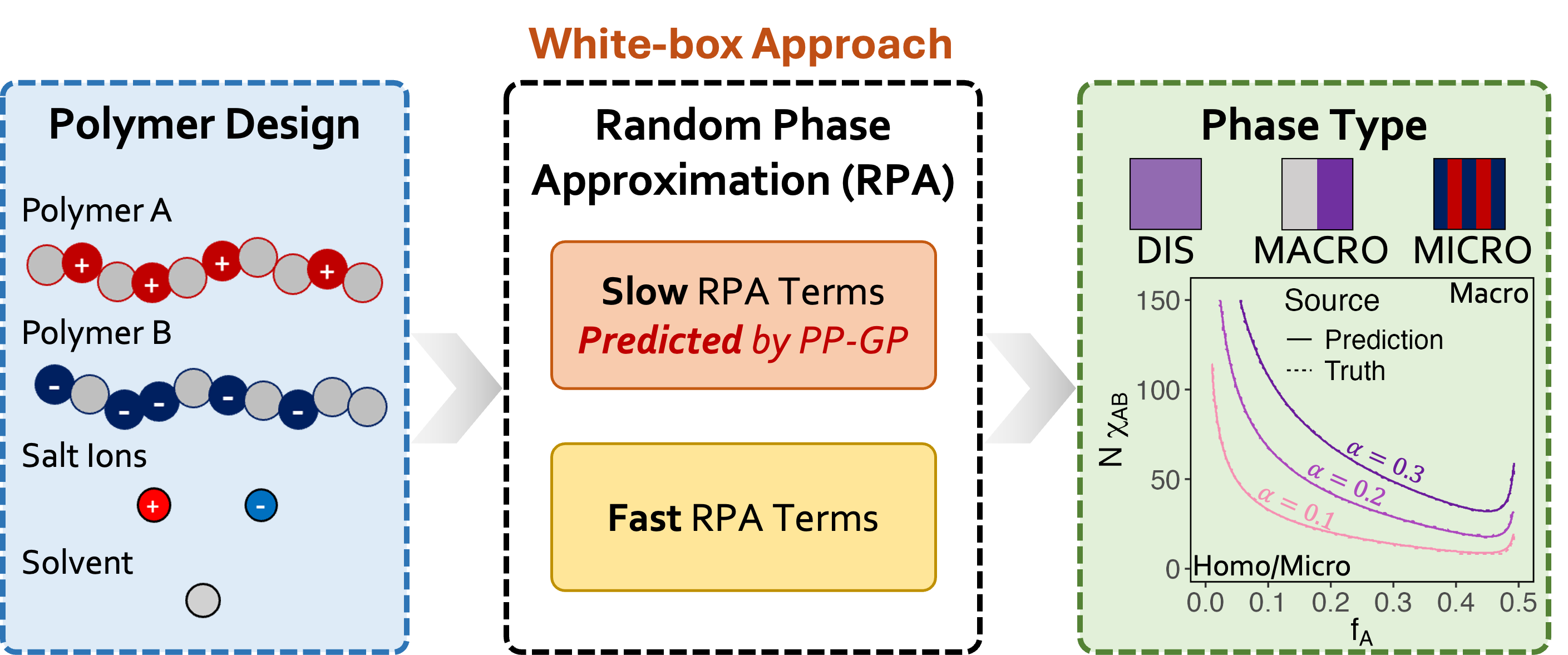}

\end{tocentry}

\newpage
\begin{abstract}

Compatibilized polymer blends are a complex, yet versatile and widespread category of material. When the components of a binary blend are immiscible, they are typically driven towards a macrophase-separated state, but with the introduction of electrostatic interactions,  they can be either homogenized or shifted to micro-phase separation. 
However, both experimental and simulation approaches face significant challenges in efficiently exploring the vast design space of charge-compatibilized polymer blends, encompassing chemical interactions, architectural properties, and composition. In this work, we introduce a white-box machine learning approach integrated with polymer field theory to predict the phase behavior of these systems, which is significantly more accurate than conventional black-box machine learning approaches. The random phase approximation (RPA) calculation is used as a testbed to determine polymer phases. Instead of directly predicting the polymer phase output of RPA calculations from a large input space by a machine learning model, we build a parallel partial Gaussian process model to predict the most computationally intensive component of the RPA calculation that only involves polymer architecture parameters as inputs. This approach substantially reduces the computational cost of the RPA calculation across a vast input space with nearly 100\% accuracy for out-of-sample prediction, enabling rapid screening of polymer blend charge-compatibilization designs. More broadly, the white-box machine learning strategy offers a promising approach for dramatic acceleration of polymer field-theoretic methods for mapping out polymer phase behavior.     
\end{abstract}


\section{Introduction}\label{sec:Intro}

The compatibilization of chemically diverse polymer blends has become an increasingly important problem to solve due to the pressing need for plastic recycling\cite{jung2023review}.  A primary challenge is that even highly chemically-similar polymers will macrophase-separate upon blending due to the very weak entropy of mixing that scales as $1/N$ where $N$ is the degree of polymerization of a given polymer\cite{flory1942thermodynamics}.
One compatibilization strategy that shows great promise is leveraging electrostatic interactions between blend components to suppress macrophase separation\cite{grzetic2021electrostatic,fredrickson2022ionic,xie2023compatibilization,edmund2025compatibilization}. By installing complementary, charged functional groups along the polymer backbone, various immiscible polymer mixtures have been rendered miscible.\cite{eisenberg1982compatibilization,russell1988microstructure,zhang2014supramolecular}  Specifically, when opposite charges are installed on dissimilar blend components, the long-range attractive electrostatic interactions in the low dielectric polymer environment compete with unfavorable mixing enthalpies to suppress macrophase separation and stabilize microphases. The resulting electrostatically-stabilized, nanostructured alloys are macroscopically homogeneous/disordered and tend to have far superior mechanical properties than unfunctionalized, phase-separated materials\cite{beech2025electrostatic}.  
However, a central challenge in this newly-proposed strategy lies in efficiently screening the large parameter space of blend designs, including monomer choice, polymer molecular weight and architecture, blend composition, and charge placement, among other variables, to obtain tailored chemical or physical properties. 


To address this challenge, machine learning  (ML) approaches are increasingly employed 
to predict material properties\cite{mysona2024machine} and accelerate material design\cite{gao2024machine}. 
A critical aspect of the polymer compatabilization screening process involves 
determining the phase behavior and miscibility of copolymer blends, since understanding the boundaries between a disordered, microphase separated, and macrophase separated system is essential for identifying optimal parameter regions for designing compatible polymers. 
Recent advances include training ML models by   chemical features\cite{arora2021random,ethier2022predicting,ethier2023integrating,ethier2024predicting} or means of experimental data sets involving small-angle X-ray scattering (SAXS) curves\cite{fang2025universal} and applying them to predict specific microphase or macrophase boundaries. ML approaches can also enhance experimental analysis of polymer assemblies by reconstructing 3D structures from scattering data\cite{wessels2021computational,heil2022computational,heil2023computational}. 

On the other hand, theory and simulation approaches, such as self-consistent field theory (SCFT) and atomistic or coarse-grained
molecular simulations, including molecular dynamics and Monte Carlo simulation, \cite{helfand1975block,leibler1980theory,matsen1994stable,vorselaars2011self}
can be used to guide the polymer design. Nonetheless, the computational cost of screening a large parameter space using any of these methods can be prohibitive.  
ML approaches have also enabled 
rapid screening of polymer compatibility based on molecular dynamics (MD) simulation\cite{xie2022accelerating} for predicting physical properties such as ionic conductivity. These black-box approaches aim to directly predict the map between the input-output pairs of the entire simulation or experimental outcomes, by training an ML surrogate based on training inputs that fill the entire input space. While the use of such ML methods is oftentimes a computationally superior way to gain insight into physical systems compared to slow simulations/calculations, 
 black-box ML models  face two key limitations. 
First, they often require large amounts of training data for problems with a large input space, and they can  be inaccurate for extrapolation \cite{ethier2025limitations}. Making predictions can also be costly for an ML black-box model trained on a large amount of data. Second, a black-box model often lacks explainability, and it can be physically incoherent due to the difficulty of integrating physical constraints into the models in some scenarios.  

In this work, we develop a novel white-box ML approach, integrated with polymer field theory, that substantially improves the accuracy and computational cost for predicting the phase behavior of charged polymer blends compared to other conventional ML approaches. We use an analytical method known as the Random Phase Approximation\cite{leibler1980theory} (RPA) as a testbed that involves 13 design parameters for determining phase behavior. The RPA is a simplification of SCFT that enables the calculation of the locus of spinodal boundaries that limit the stability of a homogeneous disordered phase\cite{fredrickson2006equilibrium}. RPA identifies these boundaries as either macrophase or microphase spinodals and, while not true binodals or order-disorder boundaries, they are close approximants to these key phase diagram features.  Numerical implementation of the RPA for phase boundary prediction is very low cost compared to alternative methods such as MD simulations or full SCFT calculations. Nonetheless, it is still not scalable for a very large design space due to the curse of dimensionality. For example, a dataset involving even just 10 points for each of the 13 input parameters employed here would require 10 trillion RPA calculations. 

Here, we find that to predict the phase portrait of simulated polymer blends, including regions corresponding to a homogeneous disordered phase (DIS), coexistence of macrophases (MACRO), or microphase separation (MICRO), conventional black-box ML approaches, such as random forest classifiers \cite{breiman2001random,Liaw2002randomforest} and neural networks\cite{lecun2015deep}, require hundreds of training samples to achieve around $80\%$ accuracy of out-of-sample prediction for samples not in the training datasets. To overcome the limit of predictive accuracy, we find that the major computational expense of RPA is due to computing the form factor matrix over reciprocal space, which is only relevant to three polymer architecture parameters. Thus, we utilized the parallel partial Gaussian process \cite{gu2016parallel} from the {\tt RobstGaSP} package \cite{gu2018robustgasp} as a surrogate model for predicting the result of an RPA calculation with an increased speed by a factor of four orders of magnitude for the costly element (polymer architecture) and two for the overall simulation. This white-box ML approach only requires around 50 training samples to achieve more than $99\%$ accuracy for out-of-sample predictions, and is hundreds of times faster than direct simulation for determining the phase behavior of a given system, making the overall strategy suitable to design space exploration.

The substantial improvement in predictive accuracy by our white-box ML approach relative to black-box ML schemes is attributed to opening up the black-box of the RPA calculation by isolating its most computationally costly component and replacing it with a surrogate model. This white-box surrogate focuses on 3 computationally slow architecture parameters out of the full 13-parameter set, and learns maps from 2 input parameters to a real-valued vector output across all Fourier basis functions, which is substantially more efficient than predicting the map between the categorical polymer phases from a 13-parameter space.
The proposed  approach enables the prediction of thousands of samples to determine phase boundaries in seconds on a desktop CPU, while maintaining near 100\% predictive accuracy. This rapid screening capability can be integrated with experimental characterization tools, as some relevant parameters, such as Flory-Huggins $\chi$-values, typically require experiments to determine\cite{mark2007physical,lee2017measurement}. Furthermore, we develop a novel uncertainty measure of the prediction that can further improve the predictive accuracy by correcting physically inconsistent predictions. The general approach is extendable to accelerate expensive simulations, as the core element of the approach is to predict the computationally intensive element by a surrogate model under physical and chemical constraints. 
The data and code used in this article are
publicly available (\url{https://github.com/UncertaintyQuantification/White-Box-ML}).
\section{Methods}

\subsection{Phase Behavior from Random Phase Approximation Calculations}\label{subsec:RPA}

The underlying microscopic model that we use to evaluate phase stability consists of an incompressible mixture of two oppositely charged polymers (A and B), small salt ions ($\pm$), and explicit solvent (S). Each polymer is defined by a discrete Gaussian chain with $N_\text{p}$ beads and charge fraction $\alpha_\text{p}$, where the charged beads are evenly distributed along the polymer backbone. For simplicity, we assign each polymer to have an identical statistical segment length, $b$, which serves as the reference length scale for the model that nondimensionalizes all other lengths. Small salt ions ensure electroneutrality when there is charge asymmetry between the polymers, but these can also be included as excess salt in the mixture.  In order to enforce incompressibility and maintain continuous parameter domains, we vary the volume fraction of solvent, $\phi_\text{S}$, and use relevant fractions to determine the composition of other components such that $\phi_\text{S} + \phi_\text{A} + \phi_\text{B} + \phi_+ + \phi_- = 1$. Namely, $f_\text{A} = \phi_\text{A}/(\phi_\text{A} + \phi_\text{B})$ is the relative fraction of $A$ monomers to the total polymer volume fraction. The parameter $\phi_p = \phi_\text{A} + \phi_\text{B} + \phi_\pm^0$ is the polymer volume fraction that includes the compensating small ion fraction, $\phi_\pm^0$. The balance of the mixture is made up of excess small ions, which can be controlled via the ratio $f_p = \phi_p/(1-\phi_\text{S})$. A system of pure polymer and compensating small ions would correspond to $\phi_\text{S} =0$ and $f_p = 1$. Due to the symmetry in chain statistics and to reduce redundancy in the parameter space, we fix $N_\text{A}=100$ and vary the chain length ratio $r=N_\text{B}/N_\text{A}$ to control the chain length of polymer B. We include both non-electrostatic (i.e., contact Flory-Huggins $\chi$) and electrostatic interactions, which are detailed along with the model derivation in the Supporting Information. Electrostatic interactions consist of direct Coulomb interactions whose strength is mediated by uniform relative permittivity that enters the bare Bjerrum length $l_\text{B} = e_0^2/(4\pi \epsilon_0 \epsilon_r k_B T)$, where $e_0$ is the elementary charge, $\epsilon_0$ is the vacuum permittivity, $\epsilon_r$ is the relative permittivity, and $k_B$ is the Boltzmann constant. For low dielectric environments of typical polymers, $l_\text{B}/b$ is $\mathcal{O}(10)$. The limit of $l_\text{B}/b \to 0$ represents a system with no electrostatic interactions. We do not include explicit polarization or composition-dependent permittivity. However, we consider dielectric contrast, $\epsilon_\text{A}/\epsilon_\text{B}$, through a Born-like contribution that manifests as a short-range attraction between small ions and the high dielectric polymer\cite{grzetic2021electrostatic}. We restrict ourselves to $\epsilon_\text{A}/\epsilon_\text{B} > 1$ so that the small ions have preferential attraction to the A polymer through an effective interaction, $\chi_{\text{A}\pm} = \frac{-l_\text{B}}{2 a_\pm} \Big(\frac{\epsilon_\text{A}}{\epsilon_\text{B}} - 1\Big)$, where $a_\pm$ is the ion size. Overall, 13 parameters comprise the polymer architecture, system composition, and interaction parameters, which are listed in Table~\ref{tab:parameters}.



\begin{table}[t]
    \centering
    \begin{tabular}{c|cm{8cm}c}
        \hline
        Category & Parameter & Description & Range \\
        \hline
        Architecture &  $r=N_\text{B}/N_\text{A}$ & Chain length asymmetry ratio & $[0.5,2.0]$ \\
         & $\alpha_\text{A} = N_{\text{A},c}/N_\text{A}$ & Charge fraction on A & $[0,1]$ \\
         & $\alpha_\text{B} = N_{\text{B},c}/N_\text{B}$ & Charge fraction on B & $[0,1]$ \\
         \hline
         Composition & $f_\text{A}=\phi_\text{A}/(\phi_\text{A} + \phi_\text{B})$ & Fraction of polymer A & $[0,1]$ \\
         & $f_p = \phi_p/(1-\phi_\text{S})$ & Relative fraction of polymer & $[0,1]$ \\
         & $\phi_\text{S}$ & Solvent volume fraction & $[0,1]$ \\
         \hline
         Interaction & $N_\text{A}\chi_{\text{AB}}$ & Chemical incompatibility between A-B & $[0,100]$ \\
         & $N_\text{A}\chi_{\text{AS}}$ & Chemical incompatibility between A-Solvent & $[0,100]$ \\
         & $r_\chi = \chi_{\text{AS}}/\chi_{\text{BS}}$ & Chemical incompatibility contrast \newline \hspace*{1em} (A-Solvent to B-Solvent) & $[0,2]$ \\
         & $\epsilon_\text{A}/\epsilon_\text{B}$ & Dielectric contrast (A-B) & $[1,5]$ \\
         & $l_\text{B}/b$ & Bare Bjerrum length & $[0,10]$ \\
         & $a_+/b$ & Salt cation size & $[0.1,10]$ \\
         & $a_-/b$ & Salt anion size & $[0.1,10]$ \\
         \hline
        
    \end{tabular}
    \caption{Summary of system parameters determined from microscopic model. The provided ranges correspond to typical physical ranges that we consider in this work.}
    \label{tab:parameters}
\end{table}

To evaluate the phase behavior, we employ a field-theoretic framework with the RPA approximation\cite{fredrickson2006equilibrium}. RPA amounts to a linear stability analysis about 
DIS for weak density fluctuations, allowing determination of spinodal instabilities (for macro or micro-phase separation) through the behavior of the static structure factor matrix, $\mathbf{S}(k)$. In this framework, the inverse structure factor at a discrete Fourier basis function 
$k\in \{k_1, ..., k_m\}$, with $m=200$ being the number of discrete Fourier basis functions, is a $7\times 7$ matrix computed by
\begin{equation}
    \mathbf{S}^{-1}(k) = \mathbf{G}^{-1}(k) + \mathbf{U}(k),
    \label{eq:S_inv}
\end{equation}
where $\mathbf{G}(k)$ is the form factor matrix, and $\mathbf{U}(k)$ represents pairwise interactions such as the $\chi$-type interactions between polymer species and electrostatic interactions among charged polymer segments and salt ions (see Section S1 of the Supporting Information for further details). 
The $7\times 7$ matrix form is a result of  pairwise relationships between the seven components of the physical system: charged/uncharged beads in polymer A, charged/uncharged beads in polymer B, cationic salt ions, anionic salt ions, and solvent molecules.  Phase stability is determined by the divergence of the determinant of the structure factor $\mathbf{S}(k)$, which can be conveniently determined by location of the $k^*$th Fourier basis such that  $|\mathbf{S}^{-1}(k^*)| = 0$ with $|\cdot|$ being the determinant operator. If no such divergence occurs, i.e., $|\mathbf{S}^{-1}(k)|>0$ for all $k$, the system is stable and forms a 
DIS. 
When a divergence exists 
and occurs at the first Fourier basis, i.e., $k^* = 0$, the blend is unstable, which results in macrophase separation, whereas if $k^* > 0$, the blend is in a state of microphase separation. 

The primary computational cost in the RPA arises from evaluating 
the form factor matrix $\mathbf G(k)$ 
which has the block diagonal structure:
\begin{equation} \label{eq:G_k}
    \mathbf{G}(k) = \begin{pmatrix}
        \frac{n_\text{A}}{V}\mathbf{\Gamma}(\mathbf{x}_\text{A},k) &0 &0 &0 &0 \\
        0& \frac{n_\text{B}}{V}\mathbf{\Gamma}(\mathbf{x}_\text{B},k) &0 &0 &0 \\
        0&0 & \frac{n_+}{V} &0 &0 \\
        0& 0&0 & \frac{n_-}{V} &0 \\
        0& 0&0 &0 & \frac{n_\text{S}}{V}
    \end{pmatrix},
\end{equation}
where each $\mathbf{\Gamma}(\mathbf{x},k)$ is a $2\times 2$ block that captures intrachain correlations between charged and uncharged beads. The number densities $n_\text{A}/V,n_\text{B}/V,n_+/V,n_-/V$ and $n_\text{S}/V$ are defined in Section S1 in the Supporting Information.  Additionally, since the entries of $\bm \Gamma(\mathbf{x},k)$ encode segment-segment correlations computed from single-chain statistics, the entries are non-negative by construction and as such, the matrix is positive definite. Specifically, the blocks have the form:
\begin{align*}
\mathbf{\Gamma}(\mathbf{x},k) = \begin{pmatrix}
    G_{11}(\mathbf{x},k) & G_{12}(\mathbf{x},k) \\ G_{12}(\mathbf{x},k) & G_{22}(\mathbf{x},k)
\end{pmatrix},
\end{align*} 
with $\mathbf{x} = \mathbf{x}_\text{A}$ or $\mathbf{x} = \mathbf{x}_\text{B}$ corresponding to polymers A and B, respectively.  Here, $\mathbf{x}_\text{A} = (N_{\text{A},c},N_{\text{A},u})$ and $\mathbf{x}_\text{B} = (N_{\text{B},c},N_{\text{B},u})$, with $N_{\text{A},c} = \alpha_\text{A}N_\text{A}$ and $N_{\text{A},u} = N_\text{A}-N_{\text{A},c}$ denoting the number of charged and uncharged beads in polymer A, respectively. The parameters $N_{\text{B},c}$ and $N_{\text{B},u}$ are defined similarly for polymer B. We suppress the dependence of $\mathbf x$, $N_c$ and $N_u$ on polymers A or B when there is no confusion. 

The computationally intensive component of RPA is the evaluation of the species-specific blocks $\mathbf{\Gamma}(\mathbf{x},k)$ separately for  polymers A and B, which require double summations over all pairs of charged and/or uncharged monomers along a single chain
\begin{align}
    &G_{11}(\mathbf{x},k) = \sum_{i=1}^{N_c+N_u}\sum_{j=1}^{N_c+N_u}\Phi(k)^{|i-j|}\delta_{i,u}\delta_{j,u},\label{eq:G11}\\
    &G_{12}(\mathbf{x},k) = G_{21}(\mathbf x, k) = \sum_{i=1}^{N_c+N_u}\sum_{j=1}^{N_c+N_u}\Phi(k)^{|i-j|}\delta_{i,u}\delta_{j,c},\label{eq:G12}\\
    &G_{22}(\mathbf{x},k) = \sum_{i=1}^{N_c+N_u}\sum_{j=1}^{N_c+N_u}\Phi(k)^{|i-j|}\delta_{i,c}\delta_{j,c},\label{eq:G22}.
\end{align}
The Kronecker delta functions $\delta_{i,c}$ and $\delta_{i,u}$ contribute only if the $i$th bead 
is charged or uncharged, respectively. These summations are weighted by powers of the Gaussian linker function $\Phi(k) = \exp(-k^2b^2/6).$

\begin{figure}[t]
    \centering
    \includegraphics[width=1\linewidth]{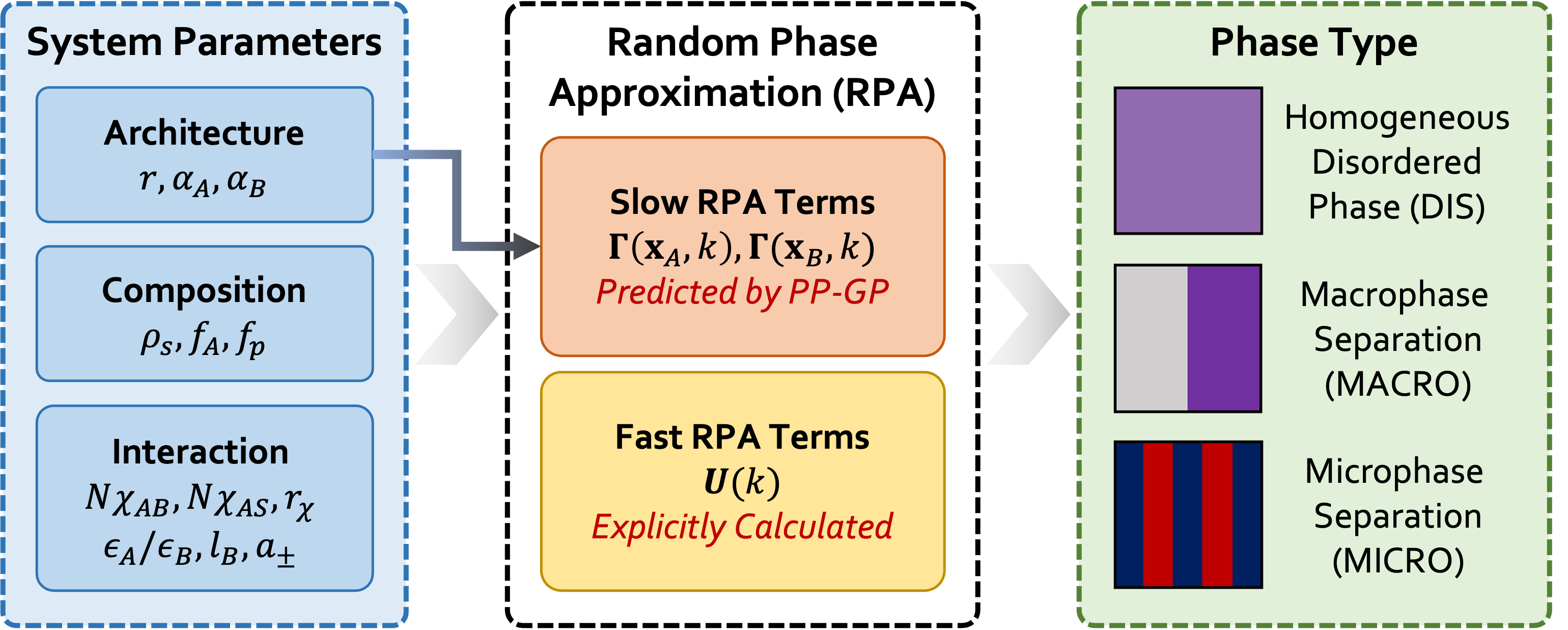}
    \caption{Workflow of random phase approximation simulation from system parameters to obtain phase behavior.}
    \label{fig:rpa_workflow}
\end{figure}

While RPA is faster than alternative approaches such as SCFT calculation or molecular dynamics simulation, directly exploring this large input space remains computationally expensive, making it challenging to design polymers of a desired phase or with tailored properties.
A conventional ML solution approach will treat RPA as a black-box function that maps from the entire parameter input space to the categorical phase output, and utilize observations that fill the entire 13 dimensional input space to train an ML classifier model. However, as the phase boundaries are complex in this large dimensional input space due to intricate interactions between parameters from Table \ref{tab:parameters}, black-box ML models require a large number of samples, making it expensive to generate accurate predictions. In seeking a way to overcome the limitation of such a conventional ML approach, we found 
that the form factor computation relevant to the 3 architecture parameters $(r,\alpha_\text{A} ,\alpha_\text{B})$ contributes the computational bottleneck in RPA, as the double summations in Equations~(\ref{eq:G11})–(\ref{eq:G22}) result in a computational cost of $\mathcal O(m(N_{c}^2+N_{u}N_{c}+N_{u}^2))$ operations over a grid of $m$ Fourier basis functions, 
which becomes prohibitive when screening a large number of systems.  Without a loss of generality, we calculate $N_{\text{A},c} = N_\text{A}\alpha_\text{A}$ and $N_{\text{A},u}=N_\text{A}-N_\text{A,c}$ to form $\mathbf x_\text{A}=(N_{\text{A},c},N_{\text{A},u})$.  For $\mathbf x_\text{B}$, after calculating $N_\text{B}=rN_\text{A}$, we calculate the number of charged/uncharged beads the same way.  For a baseline value, we use $N_\text{A}=100$. This property motivated us to develop a surrogate model to predict the expensive component of the RPA simulation, as the input space of this component only contains 3 of the 13 parameters. A workflow of RPA and our proposed white-box method for determining the polymer phase is shown in Figure
~\ref{fig:rpa_workflow}.

\subsection{Parallel Partial Gaussian Process Surrogate}\label{sec:PPGP}

To accelerate the evaluation of the form factor matrix, we construct a surrogate model using the parallel partial Gaussian process (PPGP)\cite{gu2016parallel} to replace the most computationally expensive model component of the RPA simulation, the calculation of the matrices $\bm\Gamma(\mathbf x_\text{A},k)$ and $\bm\Gamma(\mathbf x_\text{B},k)$. 
The RPA calculation requires the inverse of the form factor matrix, $\mathbf{G}^{-1}(k)$ in Equation~(\ref{eq:S_inv}), where $\mathbf{G}(k)$ is a block diagonal matrix, meaning its inverse can be computed by inverting each diagonal block independently. For the $2\times2$ block $\bm\Gamma(\mathbf x,k)$, the inverse is given by
\begin{align}
    \bm\Gamma^{-1}(\mathbf x,k) = \frac{1}{|\bm\Gamma(\mathbf x,k)|}\begin{pmatrix}
        G_{22}(\mathbf x,k) & -G_{12}(\mathbf x,k) \\
        -G_{12}(\mathbf x,k) & G_{11}(\mathbf x,k)
    \end{pmatrix},
    \label{equ:Gamma_inv}
\end{align}
where $|\bm\Gamma(\mathbf x,k)| = G_{11}(\mathbf x,k)G_{22}(\mathbf x,k)-G^2_{12}(\mathbf x,k)$. 
As the range of outcomes is large, we model the logarithm of these quantities and transform  them back for prediction, which is commonly used in surrogate models.\cite{gu2016parallel}
The surrogate takes the polymer architecture $\mathbf x=(N_c, N_u)$ as input and  $\mathbf y(\mathbf x) = [\tilde{\mathbf{G}}_{11}(\mathbf{x}),\tilde{\mathbf{G}}_{12}(\mathbf{x}),\tilde{\mathbf{G}}_{22}(\mathbf{x})]$ as output, where $\tilde{\mathbf{G}}_{g_1,g_2}(\mathbf{x}) = [\log (G_{{g_1,g_2}}(\mathbf{x},k_1)),...,\log(G_{{g_1,g_2}}(\mathbf{x},k_m))]$, for $(g_1, g_2) \in \{(1,1), (1,2), (2,2)\}$, and $\log(\cdot)$ denotes the natural logarithm operation. The full output $\mathbf y(\mathbf x)$ is therefore a vector in $\mathbb R^{3m}$, consisting of the logarithms of the required form factor entries at all $m$ wavevectors for each of the three matrix components. Here, the PP-GP surrogate is used to predict maps from $2$ dimensional inputs to $3m$ dimensional outputs: $\mathbf x \to \mathbf y(\mathbf x)$, based on a small number of training input and output pairs. 

Denote the $j$th component of $\mathbf y(\mathbf x)$ as $y_j(\mathbf x)$, for $j = 1, \dots, 3m$. For a given set of $n$ polymer architectures $\{\mathbf x_1,\dots,\mathbf x_n\}$, the vector of responses corresponding to the $j$th output coordinate, $\mathbf{y}_j = (y_j(\mathbf{x}_1),...,y_j(\mathbf{x}_n))^T$, is modeled as a multivariate normal distribution $\mathbf{y}_j \sim \mathcal{MN}(\bm{\mu}_j, \sigma_j^2(\mathbf{R} + \eta\mathbf{I}_n))$, where $\bm{\mu}_j$ is a constant mean $\bm{\mu}_j = \mathbf 1_n\mu_j$ with $\mathbf 1_n$ as an $n$-dimensional vector of ones, $\sigma_j^2$ is a variance parameter, $\eta$ is a shared nugget parameter modeling the noise, $\mathbf{I}_n$ is the $n\times n$ identity matrix and $\mathbf R$ is an $n\times n$ correlation matrix with each entry modeled by a kernel function $K(\mathbf x, \mathbf x')$ with roughness parameter $5/2$ given by,\cite{rasmussen2006gaussian} 
\begin{equation*}
    K(\mathbf{x},\mathbf{x}';\bm\gamma) = \prod_{l=1}^2 \left(1 + \sqrt{5}\frac{d_l}{\gamma_l} + \frac{5d_l^2}{3\gamma_l^2}\right)\exp\left(-\sqrt{5}\frac{d_l}{\gamma_l}\right),
\end{equation*}
where $\mathbf x=(x_1,x_2)=(N_c, N_u)$ 
$d_l = |x_l-x_l'|$ and $\gamma_l$ is the $l$th range parameter estimated by the training data, for $l=1,2$.

Given range parameters $\bm\gamma = (\gamma_1, \gamma_2)$ and nugget $\eta$, the parameters $\mu_j$ and $\sigma_j^2$ are estimated by maximizing the likelihood function to obtain $\hat{\mu}_j= (\mathbf 1_n^T(\mathbf{R}+\eta\mathbf I_n)^{-1}\mathbf 1_n)^{-1}\bm1_n^T(\mathbf{R}+\eta\mathbf I_n)^{-1}\mathbf{y}_j$ and $\hat{\sigma}^2_j=(\mathbf{y}_j-\mathbf 1_n\hat{\mu}_j)^T(\mathbf{R}+\eta\mathbf I_n)^{-1}(\mathbf{y}_j-\mathbf 1_n\hat{\mu}_j)/n$ as in a Gaussian process with scalar-valued outputs  \cite{Gu2018robustness}. The parameters $\bm\gamma$ and $\eta$ are estimated by numerically maximizing the profile likelihood with details given in Section S2.2 of the Supporting Information. 

Conditional on all parameters and the output at the $j$th dimension $\mathbf y_j$, the predictive distribution of $y_j(\mathbf x^*)$ at any $\mathbf x^*$ and coordinate $j=1,...,3m$ follows a normal distribution
\begin{align}y_j(\mathbf{x}^*)\mid\mathbf{y}_j,\hat{\mu}_j,\hat{\sigma}_j^2,\hat{\bm\gamma},\hat{\eta}\sim\mathcal{N}(\hat{y}_j(\mathbf{x}^*),\hat{\sigma}_j^2K^*),  \label{eq:pred_dist}
\end{align}
where the predictive mean and variance follow
\begin{align}
    \hat{y}_j(\mathbf{x}^*) & = \hat{\mu}_j + \mathbf{r}^T(\mathbf{x}^*)(\mathbf{R}+\eta\mathbf I_n)^{-1}(\mathbf{y}_j-\bm1_n\hat{\mu}_j) \label{eq:y_hat_j} \\
    \hat{\sigma}_j^2K^* &= \hat{\sigma}_j^2\left\{1 +\hat{\eta}- \mathbf{r}^T(\mathbf{x}^*)(\mathbf{R}+\eta\mathbf I_n)^{-1}\mathbf{r}(\mathbf{x}^*)\right\}\label{eq:K_star}
\end{align}
with $\mathbf{r}(\mathbf{x}^*) = [K(\mathbf{x}_1,\mathbf{x}^*;\hat{\bm\gamma}),...,K(\mathbf{x}_n,\mathbf{x}^*;\hat{\bm\gamma})]^T$, and $\hat{\bm \gamma}$ and $\hat \eta$ are the estimated range and nugget parameters, respectively. 

The predictive mean of PP-GP $\hat y_j(\mathbf x^*)$ is used as the prediction, and we transform it back to $\exp(\hat y_j(\mathbf x^*))$ to obtain the prediction $\hat G_{g_1,g_2}(\mathbf x^*,k_j)$ for any $g_1,g_2$, $\mathbf x^*$, and $k_j$. 
We then proceed with the same computational steps used in the RPA simulations, replacing the direct computation of $\bm\Gamma(\mathbf{x}^*,k)$ with predicted values of $\hat G_{11}(\mathbf{x}^*,k)$, $\hat G_{12}(\mathbf{x}^*,k)$, and $\hat G_{22}(\mathbf{x}^*,k)$ shown in Figure~\ref{fig:rpa_workflow}. Furthermore, the uncertainty of the prediction can be obtained based on the predictive credible interval from Equation (\ref{eq:pred_dist}). 

When using the surrogate model to predict entries in the form factor matrix, we must guarantee that these predictions result in physically valid outcomes. In particular, $\bm\Gamma(\mathbf{x},k)$ should be positive definite. However, for a small set of test inputs, we find that the predicted determinant ${| \hat{\bm\Gamma}(\mathbf{x}^*,k)|} = \hat G_{11}(\mathbf{x}^*,k)\hat G_{22}(\mathbf{x}^*,k)-(\hat G_{12}(\mathbf{x}^*,k))^2 < 0$ for some $k$, which is not physically coherent. To address this issue, we draw samples from the predictive distribution (\ref{eq:pred_dist}) and retain those that yield valid positive-definite matrices across all $3m$ grids. The final predicted phase is selected as the majority vote among the valid samples. The predictive samples can be used as a measure to determine the uncertainty of the prediction and achieve further improvement of accuracy, which will be discussed in detail later.
More details of the predictive sampling are provided in Section S2.3 of the Supporting Information.

\section{Tasks and Data}\label{sec:TaskAndData}

Our goal is to predict the phase class of polymer blends from RPA simulations at very low computational expense. 
Instead of treating phase prediction as a classification problem with 13 input variables, we develop the white-box surrogate model to predict the most computationally intensive components that map polymer architecture $(N_c, N_u)$ to the continuous $\bm\Gamma(\mathbf x, k)$ blocks in the form factor, 
which are then used within the standard RPA simulation to determine the phase. 

To generate training data, we utilize a boundary-driven maximin design set of training inputs \cite{santner2003design} over the domain of $\mathcal X = \{(N_c, N_u)\mid 50 \leq N_c+N_u\leq 200, N_c \geq 1, N_u\geq 1\}$. The lower and upper bounds of 50 and 200 for $N_c+N_u$ correspond to values of $N_\text{B}$ when $r$ is at either end of its feasible region $[0.5,2.0]$. The cases for $N_u = 0$ or $N_c = 0$ (polymer is either fully charged or fully uncharged) are excluded for prediction due to two reasons. First, these cases are fast to simulate, incurring negligible computational cost, as only a single entry in $\bm\Gamma(\mathbf x, k)$ is nonzero across all $k$. Second, the function is less smooth near these boundaries, which can degrade surrogate model performance. Details of the space-filling design are provided in Section S2.4 of the Supporting Information.


\section{Numerical Results}\label{sec:Results}
\subsection{Form Factor Prediction} \label{sec:FormFactorNumerics}
We first evaluate the predictive performance of the PPGP surrogate model on the individual entries of the $\bm\Gamma$ matrix, which forms the basis of accurate phase prediction shown in the subsequent section. 
Performance is assessed using the normalized root mean squared error (NRMSE) for each matrix entry over $n^*$ hold-out test samples $\{\mathbf{x}_1^*,...,\mathbf{x}_{n^*}^*\}$ and all $m$ grid points:
\begin{align}
    \mbox{NRMSE}_{g_1,g_2} &= \left(\frac{\sum_{i=1}^{n^*}\sum_{j=1}^{m}(G_{g_1,g_2}(\mathbf{x}^*_i,k_j)-\hat{G}_{g_1,g_2}(\mathbf{x}^*_i,k_j))^2/(n^*m)}{\sum_{i=1}^{n^*}\sum_{j=1}^{m} (G_{g_1,g_2}(\mathbf{x}^*_i,k_j) - \bar{G}_{g_1,g_2})^2/(n^*m-1)}\right)^{\frac{1}{2}}, \label{eq:NRMSE}
\end{align}
for $(g_1, g_2) \in \{(1,1), (1,2), (2,2)\}$, where $\bar{G}_{g_1,g_2} = \sum_{i=1}^{n^*}\sum_{j=1}^mG_{g_1,g_2}(\mathbf{x}^*_i,k_j)/(n^*m)$ is the sample mean of $G_{g_1,g_2}$ among the test observations $\mathbf{x}_1^*,...,\mathbf{x}_{n^*}^*$. 
A preferable model should have a smaller value of $\mbox{NRMSE}$, which indicates lower predictive error.   

\begin{figure}[t]
    \centering
    \includegraphics[width=.85\linewidth]{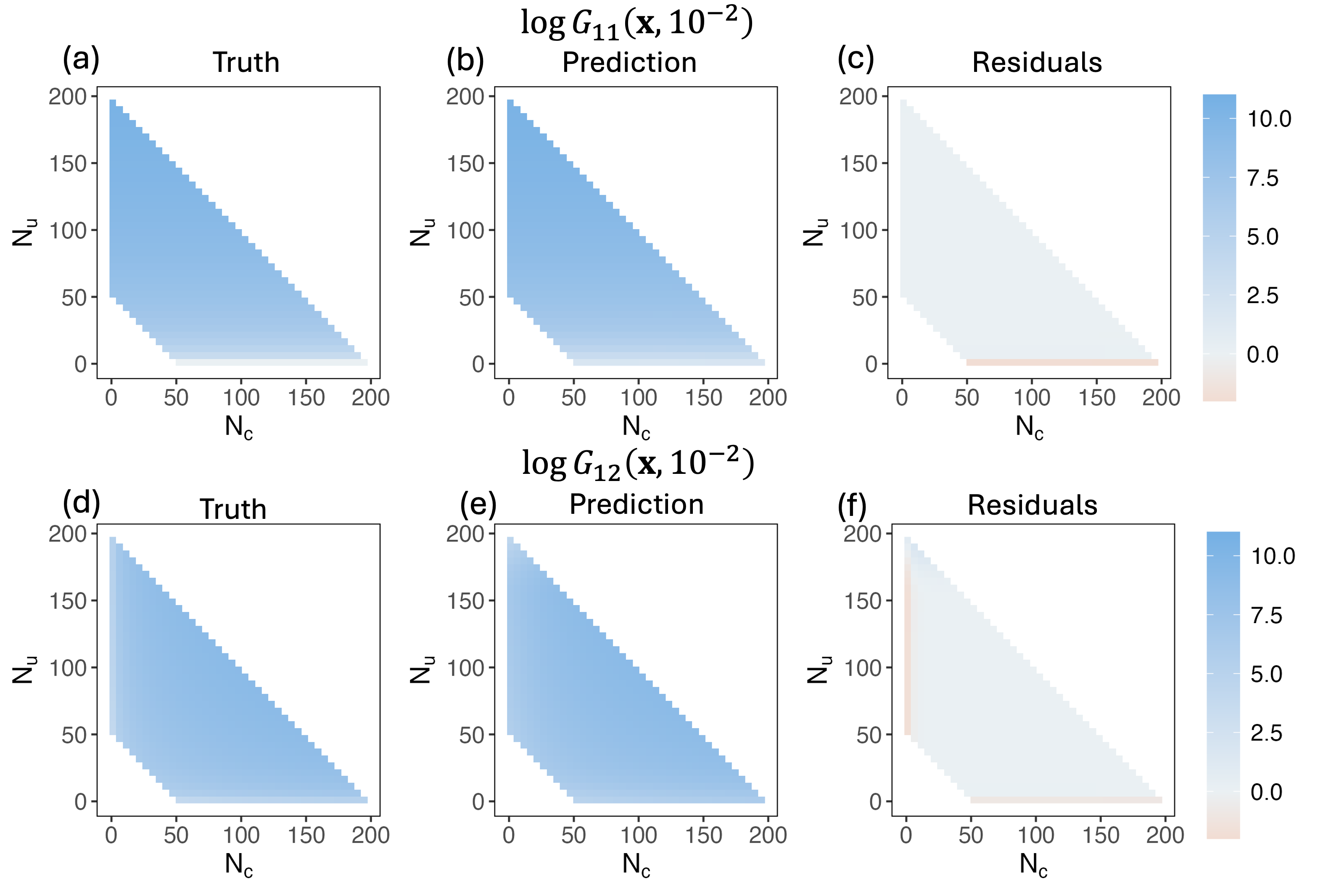}
    \caption{True, predicted and residual values of $\log(G_{11}(\mathbf{x},10^{-2}))$ (panels (a)-(c)) and $\log(G_{12}(\mathbf{x},10^{-2}))$ (panels (d) - (f)) using $n=50$ training points.  
    }
\label{fig:FormFactorPreds}
\end{figure}

We use $n=50$ training samples and evaluate predictions over the full domain of $\mathcal{X} = \{(N_c, N_u)\mid 50 \leq N_c + N_u \leq 200,\ N_c \geq 1,\ N_u \geq 1\}$, which consists of 725 hold-out test points on an equally spaced grid.  Each test input/output pair consists of $\mathbf{x}=(N_c,N_u)$ with the corresponding output $\mathbf{y}(\mathbf{x})$.  The resulting NRMSE values (Equation~(\ref{eq:NRMSE})) for each matrix entry are: $\mbox{NRMSE}_{1,1} = 0.02,\mbox{NRMSE}_{1,2}=0.068$ and $\mbox{NRMSE}_{2,2}=0.01$. All values are small, indicating accurate predictions across the domain. The cross-term $G_{12}$ exhibits slightly larger NRMSE due to its sensitivity to the lack of smoothness near the boundary as shown in panel (f) of Figure \ref{fig:FormFactorPreds}. Figure \ref{fig:FormFactorPreds} shows the predicted and true values of $\log G_{11}$ and $\log G_{12}$ at $k=0.01$, along with the corresponding residuals. We show the results in the predicted log space, rather than the original scale, to enhance visual clarity. The predictions align closely with the ground truth across the domain, with slightly larger discrepancies near the boundaries due to fewer training data. In Section S4, we show additional results for uncertainty quantification of the predictions and  NRMSEs as we increase the amount of training data in our predictions. 

\subsection{Comparison with Black-box Models for Phase Prediction}\label{sec:MethodComparison}

To evaluate the effectiveness of our white-box PPGP-surrogate approach, we compare its predictive performance with three conventional black-box machine learning methods that directly map the full set of 13 system parameters to phase labels without utilizing the underlying physical structure as follows:
\begin{itemize}
\item \textbf{NN}: A neural network (NN) classifier with three hidden layers of $512$ nodes each with ReLU activation and a final softmax output layer of size three, 
implemented using the \texttt{Keras} package.\cite{chollet2015keras} 
\item \textbf{RF}: A random forest (RF) classifier\cite{breiman2001random} that predicts phase based on predicted probability for each phase, implemented using the \texttt{RandomForest} package \cite{Liaw2002randomforest}. 
\item \textbf{GB}:  A gradient boosting (GB) model using the XGBoost algorithm \cite{chen2016xgboost} based on predictive probabilities,  
implemented using the \texttt{xgboost} package \cite{chen2025xgboostpackage}.
\end{itemize}
Implementation details are provided in Section S3 of the Supporting Information.  

We compare the out-of-sample predictive accuracy of all methods using training sizes ranging from 10 to 500 by sampling from all 13 system parameters (Table~\ref{tab:parameters}). Two test datasets are generated for evaluation. The first contains 2000 independently sampled points uniformly distributed over the full parameter space. The second test set is constructed from a $10 \times 10 \times 10$ grid over $(r, \alpha_\text{A}, \alpha_\text{B})$ and for each element of this grid, we have ten points with the remaining parameters sampled uniformly 
for a total of 10{,}000 points. For each training size, the procedure is repeated five times with different training sets generated using the space-filling method described in Section S2.4 in the Supporting Information to compute the mean and the range of classification accuracy. 

\begin{figure}[t] 
    \centering
    \includegraphics[width=\linewidth]{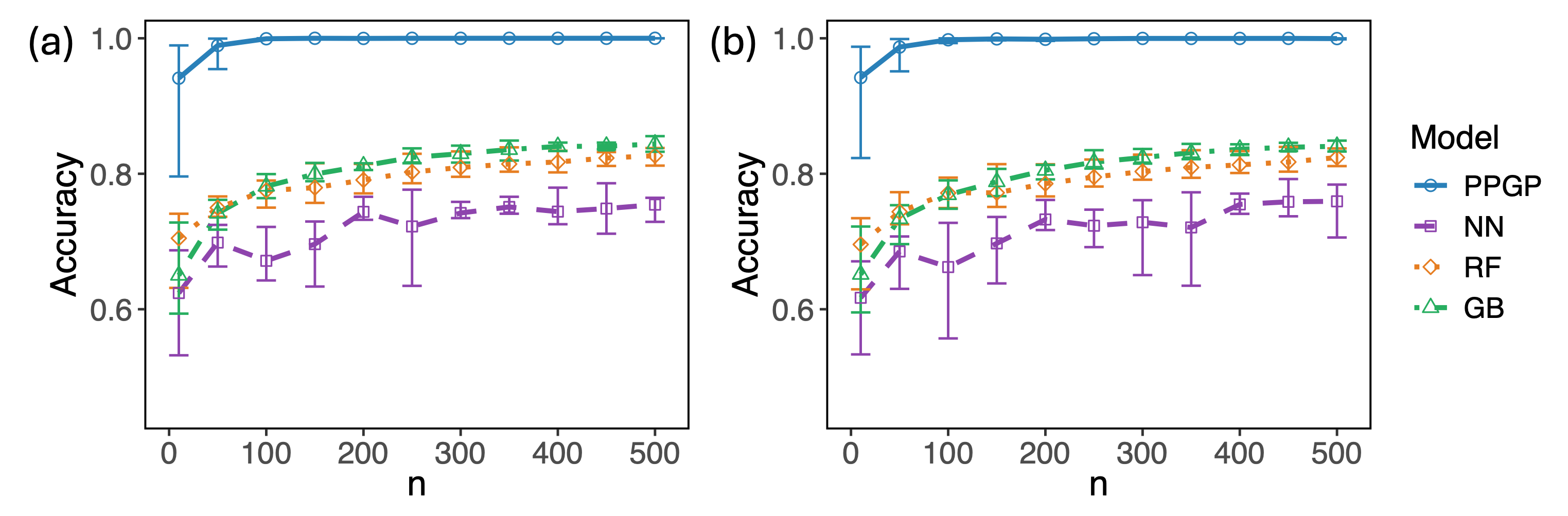}
    \caption{
    Phase prediction accuracy of four models for varying training sizes on the uniformly sampled test set (panel (a)) and the grid-based architecture test set (panel (b)). Results are averaged over predictions from models built from five different training datasets sampled according to our boundary-driven maximin design. Vertical bars indicate the minimum and maximum accuracies on the two test sets across each of these five training datasets. 
    }
    \label{fig:accuracy_w_others}
\end{figure}

The out-of-sample predictive accuracy for polymer phase prediction by different approaches are shown in Figure~\ref{fig:accuracy_w_others}. Across both test sets, our white-box method significantly outperforms all black-box alternatives, especially at small training sizes. With just 50 training points, our method already achieves greater than 99\% accuracy, whereas the best-performing black-box model (XGBoost) reaches only around 85\% with 500 training points. Moreover, we found that the variation in accuracy becomes smaller for larger training size for our method and the tree-based models, whereas the neural network exhibits larger fluctuations even with more data. This indicates our PPGP model and tree-based models are more robust in terms of variations in the training data set. 

\begin{figure}[t]
    \centering
    \includegraphics[width=0.9\linewidth]{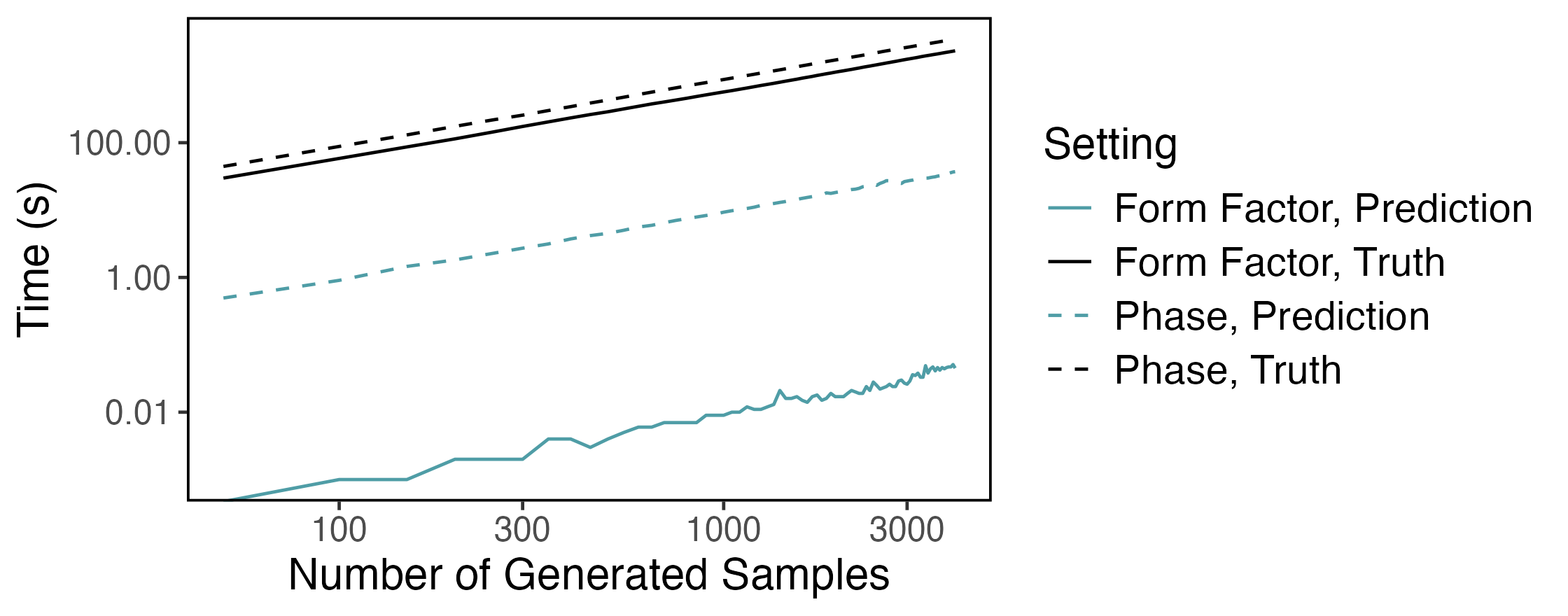}
    \caption{Computational time in seconds between the PPGP prediction using 50 training samples and full RPA simulation for obtaining $\bm{\Gamma}^{-1}(\mathbf{x},k)$ (solid lines) and phase labels (dashed lines) to generate a certain number of samples. The cost of using PPGP prediction to obtain the phase includes training PPGP models, their predictions, and all other components by plugging the PPGP predictions into RPA.
    }
    \label{fig:TimeComparison}
\end{figure}

We also compare the computational cost between our PPGP model and the full RPA simulation in Figure~\ref{fig:TimeComparison}, both for (1) computing only the inverse form factor matrix $\bm\Gamma^{-1}(\mathbf{x},k)$, and (2) running the full phase determination procedure. The surrogate model accelerates the computation of $\bm\Gamma^{-1}$ by approximately $50{,}000$-$70{,}000$ times, and reduces the overall time for phase determination by about $100$ times. While this timing study is performed for $N_\text{A}=100$, shrinking or increasing this value will speed up or slow down the simulation, respectively, whereas the cost of the prediction from our surrogate model is invariant to $N_\text{A}$ and $N_\text{B}$.  Note that the cost of training the PPGP surrogate model and making predictions is only a small fraction of the total computational cost of phase prediction, and the cost of fast-computed components in RPA now becomes the dominant cost in RPA prediction.  

\subsection{Case Studies}\label{sec:CaseStudies}
We next demonstrate the utility of our white-box surrogate model in determining polymer phases through three case studies aimed at suppressing macrophase separation in solvent-free polymer blends with no excess salt ($\phi_\text{S} = 0$ and $f_p=1$). For the purpose of compatibilization, both homogeneous phases and microphase-separated phases avoid the formation of large-scale demixing that compromises material performance. Our goal is to rapidly and accurately explore the design space to identify charge fractions that enable compatibilization between two polymers.

For each case study, we construct a $25\times 25$ grid over charge fractions $\alpha_\text{A},\alpha_\text{B} \in [0.02, 0.98]$, which ensures each polymer contains at least one charged and one uncharged bead. With these studies, we seek to show not only that our model is extremely accurate, but also that we can screen the most expensive portions of the parameter space, corresponding to the architecture parameters, with ease.  We vary one key system parameter, the chain length ratio $r$ or the chemical incompatibility between polymers A and B $N_\text{A}\chi_{\text{AB}}$, while holding all other parameters fixed at baseline values ($r=1,f_\text{A}=0.5,f_p=1,\phi_\text{S}=0,N_\text{A}\chi_{\text{AB}}=50,N_\text{A}\chi_{\text{AS}}=0,r_\chi=0,l_\text{B}/b=2,\epsilon_\text{A}/\epsilon_\text{B}=1,a_+/b=1$ and $a_-/b=1$). To assess the predictive accuracy of the model, we also run full RPA simulations for all grid points under each scenario. 



\begin{figure}[t]
    \centering
    \includegraphics[width=\linewidth]{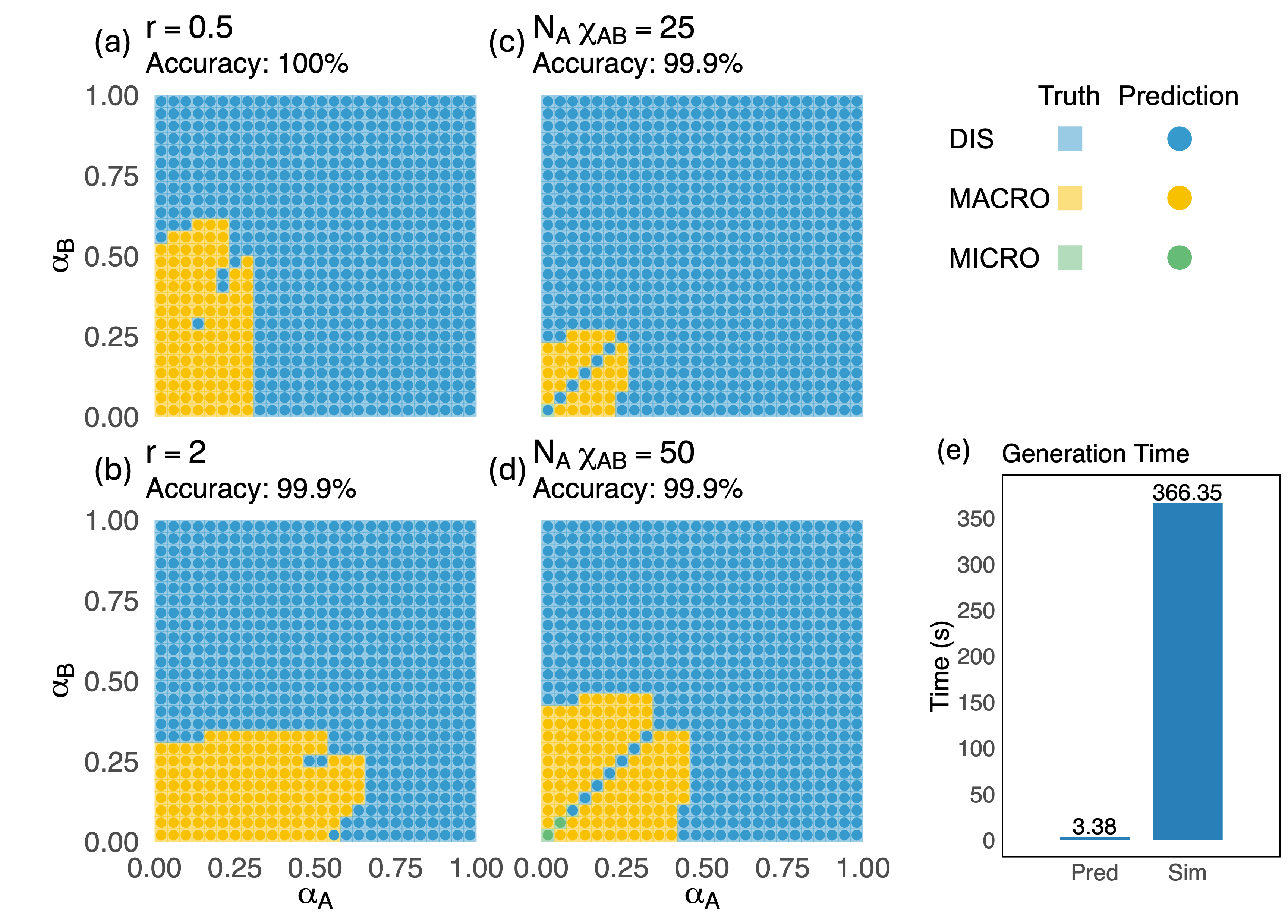}
    \caption{
    Phase diagrams in the $\alpha_\text{A}$ vs. $\alpha_\text{B}$ plane generated using the PPGP white-box method versus full RPA simulations for varying the chain length ratio $r$ (panels (a)–(b))
    and the chemical incompatibility $N_\text{A}\chi_{\text{AB}}$ (panels (c)–(d)). Panel (e) compares the computation times required to generate each diagram using direct simulation versus the PPGP surrogate model.
    }
    \label{fig:case}
\end{figure}

The results, shown in Figure~\ref{fig:case} as phase diagrams in the $\alpha_\text{A}$-$\alpha_\text{B}$ plane, indicate that our method achieves near-perfect accuracy, with at most one misclassification for each case, occurring only at phase boundaries. Each phase diagram takes just 3 seconds to generate using the surrogate, which is 100 times faster than direct simulation, which takes approximately 5 to 6 minutes, without compromising accuracy. This efficiency enables high-throughput screening for designing compatible polymers. 
Across both scenarios in Figure~\ref{fig:case}, we observe that increasing charge fractions tends to suppress macrophase separation. From panels (a) and (b) that vary the chain length ratio $r$, we observe that the shorter polymer requires a higher charge fraction to achieve compatibilization. This is because reducing the chain length while keeping the charge fraction constant decreases the total number of charged beads in the system. 
Increasing the chemical incompatibility, as shown in panels (c) and (d), expands the macrophase-separated region, which aligns with physical expectations.

\begin{figure}[t]
    \centering
    \includegraphics[width=\linewidth]{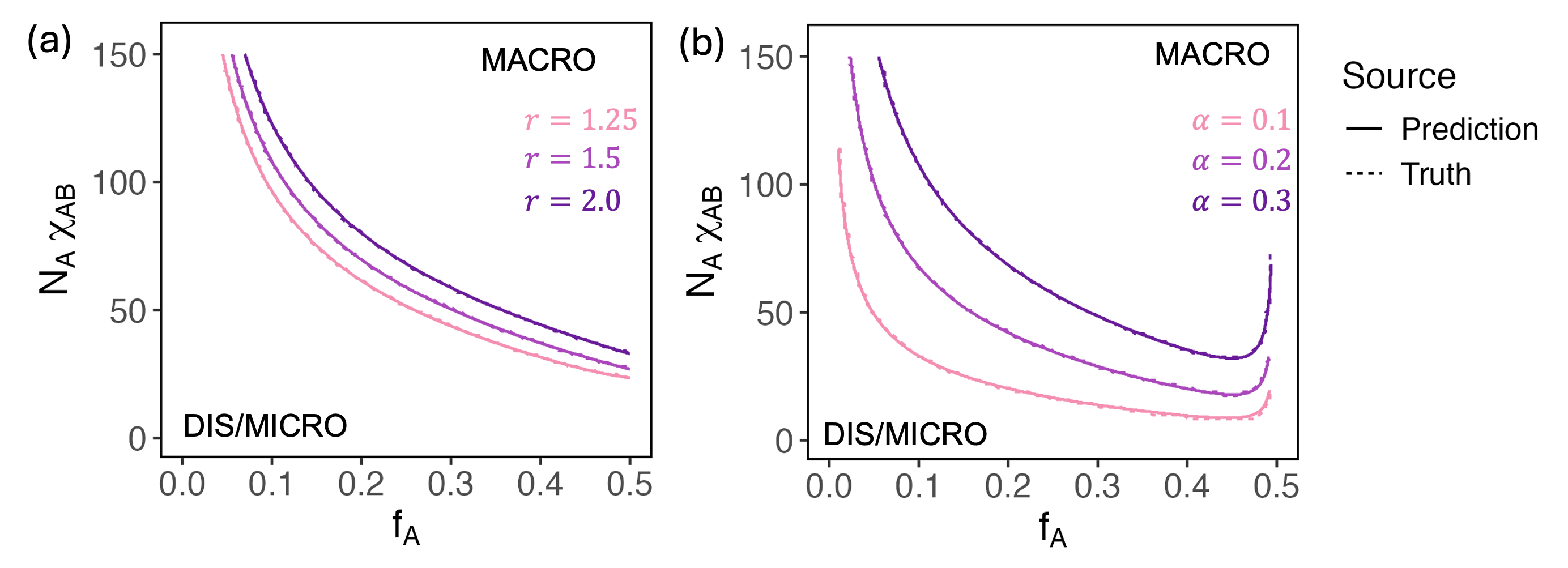}
    \caption{Phase boundaries between compatibilized (disordered or microphase-separated) and macrophase-separated states. Panel (a): critical $N_\text{A}\chi_{AB}$ as a function of $f_\text{A}$ for varying $r$. Panel (b): critical $N_\text{A}\chi_{AB}$ versus $f_\text{A}$ for different $\alpha$. 
    Unless specified otherwise, $\alpha = 0.25$ and all other parameters are set to their baseline values.  The truth and prediction lines visually overlap near-perfectly.
    }
    \label{fig:boundary_plt}
\end{figure}

Figure~\ref{fig:boundary_plt} demonstrates the use of the PPGP surrogate, trained with only $n=50$ samples, to identify phase boundaries between compatibilized states (disordered or microphase-separated) and macrophase-separated regions of the full input space under symmetric charge conditions ($\alpha_\text{A} = \alpha_\text{B} = \alpha$).  Unless specified otherwise, $\alpha=0.25$ is used as the default charge fraction.  Each panel shows the critical value of $N_\text{A}\chi_{\text{AB}}$ required to suppress macrophase separation as a function of $f_\text{A}$ or $\alpha$
, compared against results from full RPA simulations. Visually, the different lines corresponding to the predicted and true boundaries are nearly indistinguishable due to the model's near-perfect accuracy.
Across both panels, larger charge fractions $\alpha$, and compositions smaller than $f_\text{A} = 0.5$ (more asymmetric compositions) raise the threshold $N_\text{A}\chi_{\text{AB}}$ needed to prevent macrophase separation. This trend aligns with physical expectations that the incorporation of more charge or an increase in the electrostatic strength stabilizes the blend, and that increasing the volume fraction of a charged polymer adds more charge to the system, also increasing the incompatibility threshold to make the blend phase separate. The predicted boundaries closely match the simulation results, indicating the model's ability to automatically detect phase boundaries in the system. 

\subsection{Further Improvement of Accuracy}
An advantage of the PPGP surrogate is that it provides not just a point estimate but a full predictive distribution for each output.
This distribution can serve as a measure of uncertainty through the predictive variance, which can, for instance, be used to identify test samples that cannot be predicted well and control the predictive error below a specified threshold.\cite{fang2022reliable} 
In this work, we utilize the predictive distribution to obtain sample-corrected predictions:
for the small fraction of surrogate predictions 
that yield nonphysical form factor matrices $\bm\Gamma(\mathbf{x},k)$ with negative determinants, 
we draw samples from the distribution and determine the phase by majority vote among valid samples, and the results of these sample-corrected predictions are 
reported in Figures  \ref{fig:accuracy_w_others}-\ref{fig:boundary_plt}.

When the training size is small, a very small fraction of test inputs, typically less than 0.5\%, may result in no valid sampled matrices, i.e, the determinant of $ \bm\Gamma(\mathbf{x},k)$ is negative for some $k$ across all samples. In addition, certain inputs may yield two or more predicted phases among the valid samples. 
We label both situations as uncertain predictions, and applying RPA simulation for this extremely small number of samples can further improve the prediction accuracy. We refer to this strategy as prediction after adding uncertainty-driven simulation. 
\begin{figure}[t]
    \centering
    \includegraphics[width=\linewidth]{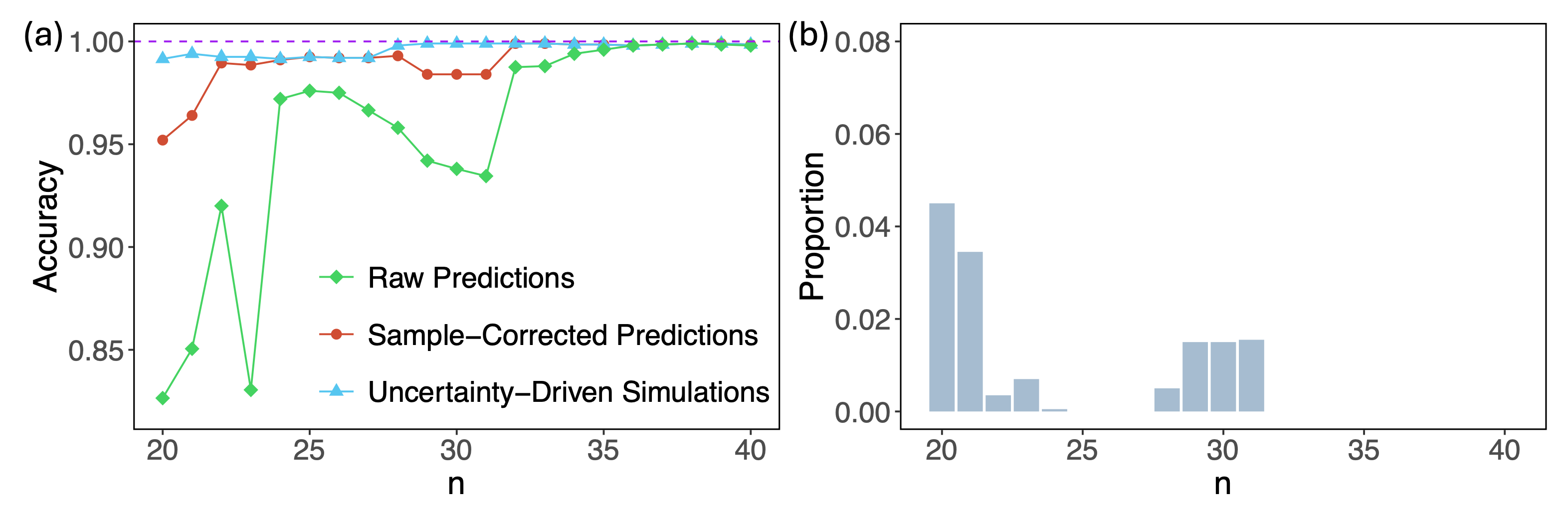}
    \caption{
    Panel (a) compares phase prediction accuracy 
    on a $2{,}000$-point test set using raw predictions 
    without correction 
    (green), sample-corrected predictions ensuring positive definite form factor matrices (red), and predictions with additional uncertainty-driven simulations 
    (blue). The dashed line indicates accuracy 1.  Panel (b) shows the proportion of test samples identified as requiring the uncertainty-driven simulation. 
    }
    \label{fig:UncertaintyPlot}
\end{figure}

Figure~\ref{fig:UncertaintyPlot}(a) presents phase prediction accuracy for three cases:  raw predictions without correction for positive definiteness, sample-corrected predictions, and predictions after adding a small set of the uncertainty-driven simulations to the predicted results, with the proportion of added simulations exactly corresponding to the test sample size shown in panel (b).
The sample-corrected predictions provide a substantial improvement over the raw predictions from 85\% to 95\% when the training size is small.
The addition of uncertainty-driven simulations offers a further improvement: with as few as 20 training points, 
accuracy rises from 95\% to over 99\%. Once the training size reaches 32, no uncertain samples remain, and the accuracy already exceeds 99\%. While our model achieves high accuracy even without adding further simulation, this internal uncertainty measure can identify almost all samples that require further inspection or simulation. 

\section{Conclusions}

In this work, we developed a white-box machine learning method to accurately classify the phase behavior of charged copolymer blends by replacing the most expensive component in polymer field theory random phase approximation calculations with a parallel partial Gaussian process surrogate model. The white-box machine learning approach 
is 4 orders of magnitude faster for form factor matrix computation and 2 orders faster for determination of polymer phases than direct simulation with RPA, enabling prediction of around 200 phases in just over 1 second on a desktop. 
The model achieves over 99\% accuracy for out-of-sample predictions of polymer phases with only 50 training samples, which is substantially more accurate than other conventional machine learning approaches considered here. 
Our approach further enables rapid screening of the full thirteen-dimensional parameter space for designing charge-compatible polymer blends.

This work opens doors for several research directions for designing polymer alloys and other formulated polymer compositions. An immediate extension is to generalize the model to include different charge distributions (e.g., block, tapered, etc.) or nonlinear chain architectures. Exploration of the expanded design space is straightforward 
and would proceed by developing appropriate training sets. The white-box surrogate strategy can also be extended to replace computationally expensive components in other simulations, such as self-consistent field theory calculations, which are significantly more costly than the random phase approximation. Furthermore, coarse-grained particle or field-based simulations of polymer blends often rely on idealized assumptions and interaction parameters that may require calibration through experiments. Integrating a machine learning surrogate to efficiently calibrate atomic-level simulations and experiments can balance the computational cost and accuracy without requiring large simulated datasets for reliable predictions. Finally, Bayesian optimization approaches that leverage the assessed uncertainty of the prediction can be used for screening  large parameter spaces to design optimal polymer formulations under structural, property, and/or cost constraints.  
\begin{acknowledgement}
This work was supported by the MRSEC Program of the National Science Foundation under Award No. DMR 2308708 (IRG-1).  Additional support for C.B. was from the National Science Foundation under the CMMT program Award No. DMR-2104255.
\end{acknowledgement}

\begin{suppinfo}

Further detail of RPA simulation; full explanation of white-box method; derivation of PPGP estimators and predictions; sampling procedure for the event the predicted $\bm\Gamma(\mathbf{x},k)$ matrices are not positive-definite; data generation process for training data; details of black-box models for comparison; additional numerical results of predicting the entries of $G_{11},G_{12},G_{22}$

\end{suppinfo}


\bibliography{References_chronical_2023}

\providecommand{\latin}[1]{#1}
\makeatletter
\providecommand{\doi}
  {\begingroup\let\do\@makeother\dospecials
  \catcode`\{=1 \catcode`\}=2 \doi@aux}
\providecommand{\doi@aux}[1]{\endgroup\texttt{#1}}
\makeatother
\providecommand*\mcitethebibliography{\thebibliography}
\csname @ifundefined\endcsname{endmcitethebibliography}
  {\let\endmcitethebibliography\endthebibliography}{}
\begin{mcitethebibliography}{42}
\providecommand*\natexlab[1]{#1}
\providecommand*\mciteSetBstSublistMode[1]{}
\providecommand*\mciteSetBstMaxWidthForm[2]{}
\providecommand*\mciteBstWouldAddEndPuncttrue
  {\def\EndOfBibitem{\unskip.}}
\providecommand*\mciteBstWouldAddEndPunctfalse
  {\let\EndOfBibitem\relax}
\providecommand*\mciteSetBstMidEndSepPunct[3]{}
\providecommand*\mciteSetBstSublistLabelBeginEnd[3]{}
\providecommand*\EndOfBibitem{}
\mciteSetBstSublistMode{f}
\mciteSetBstMaxWidthForm{subitem}{(\alph{mcitesubitemcount})}
\mciteSetBstSublistLabelBeginEnd
  {\mcitemaxwidthsubitemform\space}
  {\relax}
  {\relax}

\bibitem[Jung \latin{et~al.}(2023)Jung, Shin, Kwak, Hao, Jegal, Kim, Jeon,
  Park, and Oh]{jung2023review}
Jung,~H.; Shin,~G.; Kwak,~H.; Hao,~L.~T.; Jegal,~J.; Kim,~H.~J.; Jeon,~H.;
  Park,~J.; Oh,~D.~X. Review of polymer technologies for improving the
  recycling and upcycling efficiency of plastic waste. \emph{Chemosphere}
  \textbf{2023}, \emph{320}, 138089\relax
\mciteBstWouldAddEndPuncttrue
\mciteSetBstMidEndSepPunct{\mcitedefaultmidpunct}
{\mcitedefaultendpunct}{\mcitedefaultseppunct}\relax
\EndOfBibitem
\bibitem[Flory(1942)]{flory1942thermodynamics}
Flory,~P.~J. Thermodynamics of high polymer solutions. \emph{The Journal of
  chemical physics} \textbf{1942}, \emph{10}, 51--61\relax
\mciteBstWouldAddEndPuncttrue
\mciteSetBstMidEndSepPunct{\mcitedefaultmidpunct}
{\mcitedefaultendpunct}{\mcitedefaultseppunct}\relax
\EndOfBibitem
\bibitem[Grzetic \latin{et~al.}(2021)Grzetic, Delaney, and
  Fredrickson]{grzetic2021electrostatic}
Grzetic,~D.~J.; Delaney,~K.~T.; Fredrickson,~G.~H. Electrostatic manipulation
  of phase behavior in immiscible charged polymer blends. \emph{Macromolecules}
  \textbf{2021}, \emph{54}, 2604--2616\relax
\mciteBstWouldAddEndPuncttrue
\mciteSetBstMidEndSepPunct{\mcitedefaultmidpunct}
{\mcitedefaultendpunct}{\mcitedefaultseppunct}\relax
\EndOfBibitem
\bibitem[Fredrickson \latin{et~al.}(2022)Fredrickson, Xie, Edmund, Le, Sun,
  Grzetic, Vigil, Delaney, Chabinyc, and Segalman]{fredrickson2022ionic}
Fredrickson,~G.~H.; Xie,~S.; Edmund,~J.; Le,~M.~L.; Sun,~D.; Grzetic,~D.~J.;
  Vigil,~D.~L.; Delaney,~K.~T.; Chabinyc,~M.~L.; Segalman,~R.~A. Ionic
  compatibilization of polymers. \emph{ACS polymers Au} \textbf{2022},
  \emph{2}, 299--312\relax
\mciteBstWouldAddEndPuncttrue
\mciteSetBstMidEndSepPunct{\mcitedefaultmidpunct}
{\mcitedefaultendpunct}{\mcitedefaultseppunct}\relax
\EndOfBibitem
\bibitem[Xie \latin{et~al.}(2023)Xie, Karnaukh, Yang, Sun, Delaney, Read~de
  Alaniz, Fredrickson, and Segalman]{xie2023compatibilization}
Xie,~S.; Karnaukh,~K.~M.; Yang,~K.-C.; Sun,~D.; Delaney,~K.~T.; Read~de
  Alaniz,~J.; Fredrickson,~G.~H.; Segalman,~R.~A. Compatibilization of polymer
  blends by ionic bonding. \emph{Macromolecules} \textbf{2023}, \emph{56},
  3617--3630\relax
\mciteBstWouldAddEndPuncttrue
\mciteSetBstMidEndSepPunct{\mcitedefaultmidpunct}
{\mcitedefaultendpunct}{\mcitedefaultseppunct}\relax
\EndOfBibitem
\bibitem[Edmund \latin{et~al.}(2025)Edmund, Karnaukh, Xie, Murphy, Abdilla,
  Ino, Read~de Alaniz, Hawker, and Segalman]{edmund2025compatibilization}
Edmund,~J.; Karnaukh,~K.~M.; Xie,~S.; Murphy,~E.~A.; Abdilla,~A.; Ino,~E.;
  Read~de Alaniz,~J.; Hawker,~C.~J.; Segalman,~R.~A. Compatibilization of
  Immiscible Polymer Blends through Pendant Ionic Interactions.
  \emph{Macromolecules} \textbf{2025}, \emph{58}, 7425--7433\relax
\mciteBstWouldAddEndPuncttrue
\mciteSetBstMidEndSepPunct{\mcitedefaultmidpunct}
{\mcitedefaultendpunct}{\mcitedefaultseppunct}\relax
\EndOfBibitem
\bibitem[Eisenberg \latin{et~al.}(1982)Eisenberg, Smith, and
  Zhou]{eisenberg1982compatibilization}
Eisenberg,~A.; Smith,~P.; Zhou,~Z.-L. Compatibilization of the polystyrene/poly
  (ethyl acrylate) and polystyrene/polysoprene systems through ionic
  interactions. \emph{Polymer Engineering \& Science} \textbf{1982}, \emph{22},
  1117--1122\relax
\mciteBstWouldAddEndPuncttrue
\mciteSetBstMidEndSepPunct{\mcitedefaultmidpunct}
{\mcitedefaultendpunct}{\mcitedefaultseppunct}\relax
\EndOfBibitem
\bibitem[Russell \latin{et~al.}(1988)Russell, Jerome, Charlier, and
  Foucart]{russell1988microstructure}
Russell,~T.; Jerome,~R.; Charlier,~P.; Foucart,~M. The microstructure of block
  copolymers formed via ionic interactions. \emph{Macromolecules}
  \textbf{1988}, \emph{21}, 1709--1717\relax
\mciteBstWouldAddEndPuncttrue
\mciteSetBstMidEndSepPunct{\mcitedefaultmidpunct}
{\mcitedefaultendpunct}{\mcitedefaultseppunct}\relax
\EndOfBibitem
\bibitem[Zhang \latin{et~al.}(2014)Zhang, Kucera, Ummadisetty, Nykaza, Elabd,
  Storey, Cavicchi, and Weiss]{zhang2014supramolecular}
Zhang,~L.; Kucera,~L.~R.; Ummadisetty,~S.; Nykaza,~J.~R.; Elabd,~Y.~A.;
  Storey,~R.~F.; Cavicchi,~K.~A.; Weiss,~R. Supramolecular multiblock
  polystyrene--polyisobutylene copolymers via ionic interactions.
  \emph{Macromolecules} \textbf{2014}, \emph{47}, 4387--4396\relax
\mciteBstWouldAddEndPuncttrue
\mciteSetBstMidEndSepPunct{\mcitedefaultmidpunct}
{\mcitedefaultendpunct}{\mcitedefaultseppunct}\relax
\EndOfBibitem
\bibitem[Beech \latin{et~al.}(2025)Beech, Karnaukh, Miyamoto, Chen, Edmund,
  de~Alaniz, Hawker, and Segalman]{beech2025electrostatic}
Beech,~H.~K.; Karnaukh,~K.~M.; Miyamoto,~M.~E.; Chen,~K.; Edmund,~J.;
  de~Alaniz,~J.~R.; Hawker,~C.~J.; Segalman,~R.~A. Electrostatic
  Compatibilization of Amorphous and Semicrystalline Immiscible Polymer Blends.
  \emph{ACS Macro Letters} \textbf{2025}, \emph{14}, 969--975\relax
\mciteBstWouldAddEndPuncttrue
\mciteSetBstMidEndSepPunct{\mcitedefaultmidpunct}
{\mcitedefaultendpunct}{\mcitedefaultseppunct}\relax
\EndOfBibitem
\bibitem[Mysona \latin{et~al.}(2024)Mysona, Nealey, and
  de~Pablo]{mysona2024machine}
Mysona,~J.~A.; Nealey,~P.~F.; de~Pablo,~J.~J. Machine Learning Models and
  Dimensionality Reduction for Prediction of Polymer Properties.
  \emph{Macromolecules} \textbf{2024}, \relax
\mciteBstWouldAddEndPunctfalse
\mciteSetBstMidEndSepPunct{\mcitedefaultmidpunct}
{}{\mcitedefaultseppunct}\relax
\EndOfBibitem
\bibitem[Gao \latin{et~al.}(2024)Gao, Lin, Wang, and Du]{gao2024machine}
Gao,~L.; Lin,~J.; Wang,~L.; Du,~L. Machine learning-assisted design of advanced
  polymeric materials. \emph{Accounts of Materials Research} \textbf{2024},
  \emph{5}, 571--584\relax
\mciteBstWouldAddEndPuncttrue
\mciteSetBstMidEndSepPunct{\mcitedefaultmidpunct}
{\mcitedefaultendpunct}{\mcitedefaultseppunct}\relax
\EndOfBibitem
\bibitem[Arora \latin{et~al.}(2021)Arora, Lin, Rebello, Av-Ron, Mochigase, and
  Olsen]{arora2021random}
Arora,~A.; Lin,~T.-S.; Rebello,~N.~J.; Av-Ron,~S.~H.; Mochigase,~H.;
  Olsen,~B.~D. Random Forest Predictor for Diblock Copolymer Phase Behavior.
  \emph{ACS Macro Letters} \textbf{2021}, \emph{10}, 1339--1345\relax
\mciteBstWouldAddEndPuncttrue
\mciteSetBstMidEndSepPunct{\mcitedefaultmidpunct}
{\mcitedefaultendpunct}{\mcitedefaultseppunct}\relax
\EndOfBibitem
\bibitem[Ethier \latin{et~al.}(2022)Ethier, Casukhela, Latimer, Jacobsen,
  Rasin, Gupta, Baldwin, and Vaia]{ethier2022predicting}
Ethier,~J.~G.; Casukhela,~R.~K.; Latimer,~J.~J.; Jacobsen,~M.~D.; Rasin,~B.;
  Gupta,~M.~K.; Baldwin,~L.~A.; Vaia,~R.~A. Predicting phase behavior of linear
  polymers in solution using machine learning. \emph{Macromolecules}
  \textbf{2022}, \emph{55}, 2691--2702\relax
\mciteBstWouldAddEndPuncttrue
\mciteSetBstMidEndSepPunct{\mcitedefaultmidpunct}
{\mcitedefaultendpunct}{\mcitedefaultseppunct}\relax
\EndOfBibitem
\bibitem[Ethier \latin{et~al.}(2023)Ethier, Audus, Ryan, and
  Vaia]{ethier2023integrating}
Ethier,~J.~G.; Audus,~D.~J.; Ryan,~D.~C.; Vaia,~R.~A. Integrating theory with
  machine learning for predicting polymer solution phase behavior. \emph{Giant}
  \textbf{2023}, \emph{15}, 100171\relax
\mciteBstWouldAddEndPuncttrue
\mciteSetBstMidEndSepPunct{\mcitedefaultmidpunct}
{\mcitedefaultendpunct}{\mcitedefaultseppunct}\relax
\EndOfBibitem
\bibitem[Ethier \latin{et~al.}(2024)Ethier, Antoniuk, and
  Brettmann]{ethier2024predicting}
Ethier,~J.; Antoniuk,~E.~R.; Brettmann,~B. Predicting polymer solubility from
  phase diagrams to compatibility: a perspective on challenges and
  opportunities. \emph{Soft Matter} \textbf{2024}, \emph{20}, 5652--5669\relax
\mciteBstWouldAddEndPuncttrue
\mciteSetBstMidEndSepPunct{\mcitedefaultmidpunct}
{\mcitedefaultendpunct}{\mcitedefaultseppunct}\relax
\EndOfBibitem
\bibitem[Fang \latin{et~al.}(2025)Fang, Murphy, Kohl, Li, Hawker, Bates, and
  Gu]{fang2025universal}
Fang,~X.; Murphy,~E.~A.; Kohl,~P.~A.; Li,~Y.; Hawker,~C.~J.; Bates,~C.~M.;
  Gu,~M. Universal Phase Identification of Block Copolymers From
  Physics-Informed Machine Learning. \emph{Journal of Polymer Science}
  \textbf{2025}, \relax
\mciteBstWouldAddEndPunctfalse
\mciteSetBstMidEndSepPunct{\mcitedefaultmidpunct}
{}{\mcitedefaultseppunct}\relax
\EndOfBibitem
\bibitem[Wessels and Jayaraman(2021)Wessels, and
  Jayaraman]{wessels2021computational}
Wessels,~M.~G.; Jayaraman,~A. Computational reverse-engineering analysis of
  scattering experiments (CREASE) on amphiphilic block polymer solutions:
  cylindrical and fibrillar assembly. \emph{Macromolecules} \textbf{2021},
  \emph{54}, 783--796\relax
\mciteBstWouldAddEndPuncttrue
\mciteSetBstMidEndSepPunct{\mcitedefaultmidpunct}
{\mcitedefaultendpunct}{\mcitedefaultseppunct}\relax
\EndOfBibitem
\bibitem[Heil \latin{et~al.}(2022)Heil, Patil, Dhinojwala, and
  Jayaraman]{heil2022computational}
Heil,~C.~M.; Patil,~A.; Dhinojwala,~A.; Jayaraman,~A. Computational
  reverse-engineering analysis for scattering experiments (CREASE) with machine
  learning enhancement to determine structure of nanoparticle mixtures and
  solutions. \emph{ACS Central Science} \textbf{2022}, \emph{8},
  996--1007\relax
\mciteBstWouldAddEndPuncttrue
\mciteSetBstMidEndSepPunct{\mcitedefaultmidpunct}
{\mcitedefaultendpunct}{\mcitedefaultseppunct}\relax
\EndOfBibitem
\bibitem[Heil \latin{et~al.}(2023)Heil, Ma, Bharti, and
  Jayaraman]{heil2023computational}
Heil,~C.~M.; Ma,~Y.; Bharti,~B.; Jayaraman,~A. Computational
  reverse-engineering analysis for scattering experiments for form factor and
  structure factor determination (“p (q) and s (q) crease”). \emph{JACS Au}
  \textbf{2023}, \emph{3}, 889--904\relax
\mciteBstWouldAddEndPuncttrue
\mciteSetBstMidEndSepPunct{\mcitedefaultmidpunct}
{\mcitedefaultendpunct}{\mcitedefaultseppunct}\relax
\EndOfBibitem
\bibitem[Helfand(1975)]{helfand1975block}
Helfand,~E. Block Copolymer Theory. III. Statistical Mechanics of the
  Microdomain Structure. \emph{Macromolecules} \textbf{1975}, \emph{8},
  552--556\relax
\mciteBstWouldAddEndPuncttrue
\mciteSetBstMidEndSepPunct{\mcitedefaultmidpunct}
{\mcitedefaultendpunct}{\mcitedefaultseppunct}\relax
\EndOfBibitem
\bibitem[Leibler(1980)]{leibler1980theory}
Leibler,~L. Theory of Microphase Separation in Block Copolymers.
  \emph{Macromolecules} \textbf{1980}, \emph{13}, 1602--1617\relax
\mciteBstWouldAddEndPuncttrue
\mciteSetBstMidEndSepPunct{\mcitedefaultmidpunct}
{\mcitedefaultendpunct}{\mcitedefaultseppunct}\relax
\EndOfBibitem
\bibitem[Matsen and Schick(1994)Matsen, and Schick]{matsen1994stable}
Matsen,~M.~W.; Schick,~M. Stable and Unstable Phases of a Diblock Copolymer
  Melt. \emph{Physical Review Letters} \textbf{1994}, \emph{72}, 2660\relax
\mciteBstWouldAddEndPuncttrue
\mciteSetBstMidEndSepPunct{\mcitedefaultmidpunct}
{\mcitedefaultendpunct}{\mcitedefaultseppunct}\relax
\EndOfBibitem
\bibitem[Vorselaars \latin{et~al.}(2011)Vorselaars, Kim, Chantawansri,
  Fredrickson, and Matsen]{vorselaars2011self}
Vorselaars,~B.; Kim,~J.~U.; Chantawansri,~T.~L.; Fredrickson,~G.~H.;
  Matsen,~M.~W. Self-Consistent Field Theory for Diblock Copolymers Grafted to
  a Sphere. \emph{Soft Matter} \textbf{2011}, \emph{7}, 5128--5137\relax
\mciteBstWouldAddEndPuncttrue
\mciteSetBstMidEndSepPunct{\mcitedefaultmidpunct}
{\mcitedefaultendpunct}{\mcitedefaultseppunct}\relax
\EndOfBibitem
\bibitem[Xie \latin{et~al.}(2022)Xie, France-Lanord, Wang, Lopez, Stolberg,
  Hill, Leverick, Gomez-Bombarelli, Johnson, Shao-Horn, \latin{et~al.}
  others]{xie2022accelerating}
Xie,~T.; France-Lanord,~A.; Wang,~Y.; Lopez,~J.; Stolberg,~M.~A.; Hill,~M.;
  Leverick,~G.~M.; Gomez-Bombarelli,~R.; Johnson,~J.~A.; Shao-Horn,~Y.; others
  Accelerating amorphous polymer electrolyte screening by learning to reduce
  errors in molecular dynamics simulated properties. \emph{Nature
  communications} \textbf{2022}, \emph{13}, 3415\relax
\mciteBstWouldAddEndPuncttrue
\mciteSetBstMidEndSepPunct{\mcitedefaultmidpunct}
{\mcitedefaultendpunct}{\mcitedefaultseppunct}\relax
\EndOfBibitem
\bibitem[Ethier \latin{et~al.}(2025)Ethier, Paluch, and
  Varshney]{ethier2025limitations}
Ethier,~J.~G.; Paluch,~A.~S.; Varshney,~V. Limitations of theory-informed
  machine learning algorithms for the prediction and exploration of molecular
  properties: Solvation free energy as a case study. \emph{APL Machine
  Learning} \textbf{2025}, \emph{3}\relax
\mciteBstWouldAddEndPuncttrue
\mciteSetBstMidEndSepPunct{\mcitedefaultmidpunct}
{\mcitedefaultendpunct}{\mcitedefaultseppunct}\relax
\EndOfBibitem
\bibitem[Fredrickson(2006)]{fredrickson2006equilibrium}
Fredrickson,~G. \emph{The equilibrium theory of inhomogeneous polymers}; Oxford
  University Press, 2006\relax
\mciteBstWouldAddEndPuncttrue
\mciteSetBstMidEndSepPunct{\mcitedefaultmidpunct}
{\mcitedefaultendpunct}{\mcitedefaultseppunct}\relax
\EndOfBibitem
\bibitem[Breiman(2001)]{breiman2001random}
Breiman,~L. Random forests. \emph{Machine learning} \textbf{2001}, \emph{45},
  5--32\relax
\mciteBstWouldAddEndPuncttrue
\mciteSetBstMidEndSepPunct{\mcitedefaultmidpunct}
{\mcitedefaultendpunct}{\mcitedefaultseppunct}\relax
\EndOfBibitem
\bibitem[Liaw and Wiener(2002)Liaw, and Wiener]{Liaw2002randomforest}
Liaw,~A.; Wiener,~M. Classification and Regression by randomForest. \emph{R
  News} \textbf{2002}, \emph{2}, 18--22\relax
\mciteBstWouldAddEndPuncttrue
\mciteSetBstMidEndSepPunct{\mcitedefaultmidpunct}
{\mcitedefaultendpunct}{\mcitedefaultseppunct}\relax
\EndOfBibitem
\bibitem[LeCun \latin{et~al.}(2015)LeCun, Bengio, and Hinton]{lecun2015deep}
LeCun,~Y.; Bengio,~Y.; Hinton,~G. Deep learning. \emph{Nature} \textbf{2015},
  \emph{521}, 436--444\relax
\mciteBstWouldAddEndPuncttrue
\mciteSetBstMidEndSepPunct{\mcitedefaultmidpunct}
{\mcitedefaultendpunct}{\mcitedefaultseppunct}\relax
\EndOfBibitem
\bibitem[Gu and Berger(2016)Gu, and Berger]{gu2016parallel}
Gu,~M.; Berger,~J.~O. Parallel partial {G}aussian process emulation for
  computer models with massive output. \emph{The Annals of Applied Statistics}
  \textbf{2016}, \emph{10}, 1317--1347\relax
\mciteBstWouldAddEndPuncttrue
\mciteSetBstMidEndSepPunct{\mcitedefaultmidpunct}
{\mcitedefaultendpunct}{\mcitedefaultseppunct}\relax
\EndOfBibitem
\bibitem[Gu \latin{et~al.}(2019)Gu, Palomo, and Berger]{gu2018robustgasp}
Gu,~M.; Palomo,~J.; Berger,~J.~O. {RobustGaSP: Robust Gaussian Stochastic
  Process Emulation in R}. \emph{{The R Journal}} \textbf{2019}, \emph{11},
  112--136\relax
\mciteBstWouldAddEndPuncttrue
\mciteSetBstMidEndSepPunct{\mcitedefaultmidpunct}
{\mcitedefaultendpunct}{\mcitedefaultseppunct}\relax
\EndOfBibitem
\bibitem[Mark and Mark(2007)Mark, and Mark]{mark2007physical}
Mark,~J.~E.; Mark,~J.~E. \emph{Physical properties of polymers handbook};
  Springer, 2007; Vol. 1076\relax
\mciteBstWouldAddEndPuncttrue
\mciteSetBstMidEndSepPunct{\mcitedefaultmidpunct}
{\mcitedefaultendpunct}{\mcitedefaultseppunct}\relax
\EndOfBibitem
\bibitem[Lee \latin{et~al.}(2017)Lee, Yao, Li, Jun, and
  Lee]{lee2017measurement}
Lee,~J.~K.; Yao,~S.~X.; Li,~G.; Jun,~M.~B.; Lee,~P.~C. Measurement methods for
  solubility and diffusivity of gases and supercritical fluids in polymers and
  its applications. \emph{Polymer reviews} \textbf{2017}, \emph{57},
  695--747\relax
\mciteBstWouldAddEndPuncttrue
\mciteSetBstMidEndSepPunct{\mcitedefaultmidpunct}
{\mcitedefaultendpunct}{\mcitedefaultseppunct}\relax
\EndOfBibitem
\bibitem[Rasmussen(2006)]{rasmussen2006gaussian}
Rasmussen,~C.~E. \emph{Gaussian processes for machine learning}; MIT Press,
  2006\relax
\mciteBstWouldAddEndPuncttrue
\mciteSetBstMidEndSepPunct{\mcitedefaultmidpunct}
{\mcitedefaultendpunct}{\mcitedefaultseppunct}\relax
\EndOfBibitem
\bibitem[Gu \latin{et~al.}(2018)Gu, Wang, and Berger]{Gu2018robustness}
Gu,~M.; Wang,~X.; Berger,~J.~O. Robust {G}aussian stochastic process emulation.
  \emph{Annals of Statistics} \textbf{2018}, \emph{46}, 3038--3066\relax
\mciteBstWouldAddEndPuncttrue
\mciteSetBstMidEndSepPunct{\mcitedefaultmidpunct}
{\mcitedefaultendpunct}{\mcitedefaultseppunct}\relax
\EndOfBibitem
\bibitem[Santner \latin{et~al.}(2003)Santner, Williams, and
  Notz]{santner2003design}
Santner,~T.~J.; Williams,~B.~J.; Notz,~W.~I. \emph{The design and analysis of
  computer experiments}; Springer Science \& Business Media, 2003\relax
\mciteBstWouldAddEndPuncttrue
\mciteSetBstMidEndSepPunct{\mcitedefaultmidpunct}
{\mcitedefaultendpunct}{\mcitedefaultseppunct}\relax
\EndOfBibitem
\bibitem[Chollet \latin{et~al.}(2015)Chollet, \latin{et~al.}
  others]{chollet2015keras}
Chollet,~F.; others Keras. \url{https://keras.io}, 2015\relax
\mciteBstWouldAddEndPuncttrue
\mciteSetBstMidEndSepPunct{\mcitedefaultmidpunct}
{\mcitedefaultendpunct}{\mcitedefaultseppunct}\relax
\EndOfBibitem
\bibitem[Chen and Guestrin(2016)Chen, and Guestrin]{chen2016xgboost}
Chen,~T.; Guestrin,~C. {XGBoost: A scalable tree boosting system}. Proceedings
  of the 22nd acm sigkdd international conference on knowledge discovery and
  data mining. 2016; pp 785--794\relax
\mciteBstWouldAddEndPuncttrue
\mciteSetBstMidEndSepPunct{\mcitedefaultmidpunct}
{\mcitedefaultendpunct}{\mcitedefaultseppunct}\relax
\EndOfBibitem
\bibitem[Chen \latin{et~al.}(2025)Chen, He, Benesty, Khotilovich, Tang, Cho,
  Chen, Mitchell, Cano, Zhou, Li, Xie, Lin, Geng, Li, Yuan, and
  Cortes]{chen2025xgboostpackage}
Chen,~T. \latin{et~al.}  xgboost: Extreme Gradient Boosting. 2025; R package
  version 1.7.7.1\relax
\mciteBstWouldAddEndPuncttrue
\mciteSetBstMidEndSepPunct{\mcitedefaultmidpunct}
{\mcitedefaultendpunct}{\mcitedefaultseppunct}\relax
\EndOfBibitem
\bibitem[Fang \latin{et~al.}(2022)Fang, Gu, and Wu]{fang2022reliable}
Fang,~X.; Gu,~M.; Wu,~J. Reliable Emulation of Complex Functionals by Active
  Learning with Error Control. \emph{The Journal of Chemical Physics}
  \textbf{2022}, \emph{157}, 214109\relax
\mciteBstWouldAddEndPuncttrue
\mciteSetBstMidEndSepPunct{\mcitedefaultmidpunct}
{\mcitedefaultendpunct}{\mcitedefaultseppunct}\relax
\EndOfBibitem
\end{mcitethebibliography}

\end{document}